# Manipulating propagation and evolution of polarization singularities in composite Bessel-like fields

XINGLIN WANG,[1,4] WENXIANG YAN,[1] YUAN GAO,[1] ZHENG YUAN,[1] ZHI-CHENG REN,[1] XI-LIN WANG,[1] JIANPING DING,[1,2,3,*] AND HUI-TIAN WANG[1]

[1]Nanjing University, National Laboratory of Solid Microstructures and School of Physics, Nanjing, China

[2]Nanjing University, Collaborative Innovation Center of Advanced Microstructures, Nanjing, China

[3]Nanjing University, Collaborative Innovation Center of Solid-State Lighting and Energy-Saving Electronics, Nanjing, China

[4]Anhui Polytechnic University, Department of Applied Mathematics and Physics, Wuhu, China

*Corresponding author: jpding@nju.edu.cn

**Abstract.** Structured optical fields embedded with polarization singularities (PSs) have attracted extensive attention due to their capability to retain topological invariance during propagation. Many advances in PSs research have been made over the past 20 years in the areas of mathematical description, generation and detection technologies, propagation dynamics, and applications. However, one of the most crucial and difficult tasks continues to be manipulating PSs with multiple degrees of freedom, especially in three-dimensional (3D) tailored optical fields. We propose and demonstrate the longitudinal PS lines obtained by superimposing Bessel-like modes with orthogonal polarization states on composite vector optical fields (VOFs). The embedded PSs in the fields can be manipulated to propagate robustly along arbitrary trajectories, or to annihilate, revive, and transform each other at on-demand positions in 3D space, allowing complex PSs topological morphology and intensity pattern to be flexibly customized. Our findings could spur further research into singular optics and help with applications such as micromanipulation, microstructure fabrication, and optical encryption.

## 1. INTRODUCTION

Singularities in topological systems have been studied in a variety of disciplines, including quantum condensates [1, 2], fluid dynamics [3], superconductors [4], and field theory [5, 6]. In wave optics, singularities are introduced to describe the striking feature that parameters characterizing optical fields cannot be defined at specific positions, whereas the distribution of amplitude, phase, or state of polarization (SOP) around the respective singularity forms specific geometric configuration, namely topological morphology or skeleton of the structured optical field [7-12]. To be specific, phase singularities in scalar optical fields represent points of indeterminate phase in two-dimensional (2D) space, leading to optical vortices in spatially varying amplitude and phase structures. For vector optical fields (VOFs), polarization singularities (PSs) are used to describe the singular feature that parameters determining the local SOP cannot be defined. Generally, PSs in transverse planes are isolated C-points, points of circular polarization where the orientation of polarization ellipse is undefined, V-points, null intensity points with both indeterminate handedness and orientation of polarization ellipse, and L-lines, the handedness of polarization ellipse on them is undefined [10, 11]. In addition, the description of PSs can be conveniently extended to three-dimensional (3D) space, and then singular points and lines evolve into singular lines and surfaces [12, 13], respectively. PSs embedded in VOFs, as opposed to phase singularities, have a rich spatial distribution of SOP and topological morphology, such as lemon, star, monstar, Möbius strips, links, and knots [10, 14-17], which have been claimed to inherit persistence even when perturbed during beam propagation.

To date, significant progress has been made in PSs, including methods for generation and detection, propagation dynamics, and applications [18-20]. With increasing experimental realization and advanced technologies for VOFs, such as q-plates [21], spatial light modulators (SLMs) [22], and metasurfaces [23], complex PS configurations and PS arrays in 2D space have been successfully created [24-30], revealing some important topological properties such as sign rules, net-zero topological charge, and singularity index conservation. Whereas, as previously stated, PSs appear naturally as phenomena in 3D space, a complete picture of evolution along the third spatial dimension is especially important. PSs petal structures [31], propagation stability of on-axis PSs [21], Hilbert Hotel-like behavior in the creation of PSs [32], pseudo-topological property arising from the invisible redistribution of both spin angular momentum (SAM) and orbital angular momentum (OAM) states [33], linked and knotted longitudinal PS lines [15, 16], and Self-imaging PS networks [34] have recently been theoretically or experimentally investigated. These studies demonstrated several novel phenomena and investigated the physical implications of PSs during propagation. However, due to the topological complexity of PSs and difficulties in the experimental implementation of diverse structured optical beams, research on actively manipulating the propagation and evolution of PSs, as well as customizing on-demand 3D PSs topological configuration, remains extremely scarce and challenging.

In this paper, we employ non-diffraction Bessel beams as orthogonal components to compose Bessel-like VOFs embedded with PSs using our experiment setup, which consists of a VOF generator and a far-field measurement system. Leveraging our angular spectrum design proposed in Ref. [35], we dynamically regulate the propagation and evolution of the PSs by refreshing the computer-generated holograms (CGHs) encoded on a phase-only SLM, allowing us to easily customize complex PS topological configuration and intensity pattern in 3D space.

## 2. THEORETICAL BASIS

To obtain the desired VOFs and manipulate the embedded PSs, we resort to Bessel beams acting as orthogonal components because of their diffraction-resilience. Furthermore, such beams can be engineered to propagate with a tunable axial intensity within a range of accessible spatial frequencies [36, 37], or along curved propagation trajectories, resulting in the self-accelerating Bessel-like beams [38, 39]. To further enhance the applicability, we recently proposed an approach to generate self-accelerating zeroth-order Bessel beams with on-demand tailored intensity profiles along arbitrary trajectories, and Bessel-Poincaré beams propagating with embedded radially self-accelerating Stokes vortices [35, 40], which have the z-axial amplitude distribution of $U(z, r=0) = \sqrt{I(z)} \exp(ik_{z0}z)$ in the cylindrical coordinates $(r, \varphi, z)$, and can be calculated from the angular spectrum of the beam given by

$$A(\sqrt{k^2 - k_z^2}) = \frac{2}{rect\left(\frac{k_z}{2k}\right)} k_z \int_{-\infty}^{\infty} \sqrt{I(z)} e^{ik_{z0}z} e^{-ik_z z} dz \qquad (1)$$

where $k = 2\pi/\lambda$ is the wavevector with $\lambda$ being the wavelength of the monochromatic light, $k_z$ and $k_r$ are the longitudinal and radially wavevectors connected by $k_z = \sqrt{k^2 - k_r^2}$, $\sqrt{I(z)}$ corresponds to the z-axial intensity distribution, and rect(.) represents the rectangle function. According to the Fourier phase-shifting theorem, a light sheet within a tiny longitudinal region $(z, z+\Delta z)$ will undergo a translation from $(0,0,z)$ to $(h(z), g(z), z)$ if a complex exponential $\exp(ik_x h(z) + ik_y g(z))$ is imposed on the angular spectrum. Here we extend such a spatial spectrum method to a higher-order Bessel beam design, and the corresponding angular spectrum formula in Ref. [35] should be rewritten as

$$A_i(\sqrt{k^2 - k_{zi}^2}) = \frac{2}{rect\left(\frac{k_{zi}}{2k}\right)} k_{zi} \int_{-\infty}^{\infty} \sqrt{I_i(z)} e^{im_i \varphi} e^{i[k_{z0}z + k_{xi}h_i(z) + k_{yi}g_i(z)]} e^{-ik_{zi}z} dz \quad (2)$$

where the subcript $i = R$ and $L$ denotes the respective right-handed (RH) and left-handed (LH) circular polarization state, resecptively, and $m_i$ is the topological charge of the component phase vortex. Eq. (2) provides a feasible scheme for controlling both the intensity and propagation trajectory of higher-order Bessel modes, allowing for manipulation of the propagation and evolution of the PSs embedded in the composite Bessel-like field, as well as customization of the PS topological morphology. By using the inverse Fourier transform of Eq. (2), we can obtain the two orthogonal constituent Bessel-like beams, i.e. $E_L$ and $E_R$, needed to compose the desired VOFs. Based on the Stokes polarimetry method, the polarization distribution and PS topological morphology of the field can be determined by measuring four Stokes parameters, which are constructed in terms of circularly polarized base vector as

$$\begin{aligned} S_0 &= |E_L|^2 + |E_R|^2 \\ S_1 &= 2\operatorname{Re}(E_R^* E_L) \\ S_2 &= 2\operatorname{Im}(E_R^* E_L) \\ S_3 &= |E_L|^2 - |E_R|^2 \end{aligned} \quad (3)$$

Then local SOP of any optical field can be depicted by a polarization ellipse with the ellipticity and azimuthal angles expressed as $\chi = \arcsin(S_3/S_0)/2$ and $\psi = \arctan(S_2/S_1)/2$, respectively. Generally, PSs are examined in transverse planes, e.g. C-points and V-points, and can be identified by constructing the Stokes complex field $S_{12} = S_1 + iS_2$ with phase distribution $\phi_{12} = \arctan(S_2/S_1)$. Because C- and V-points appear in VOFs as phase vortices in the Stokes field $S_{12}$, the singular indices characterizing the polarization variation surrounding PSs can be usefully subsumed under a single integer Stokes index, defined by $\sigma_{12} = (\Delta\phi_{12})/2\pi$ with $\Delta\phi_{12}$ being the accumulated phase acquired around a closed path embracing the $S_{12}$ phase vortices in the anti-clock sense. The C-point index ($I_C$) and V-point index ($\eta$, also called Poincaré-Hopf index) connected to the Stokes index are described as $I_C = \sigma_{12}/2$ and $\eta = \sigma_{12}/2$, respectively. The polarization distribution surrounding low-order V-points, i.e. $\eta = \pm 1$ ($\sigma_{12} = \pm 2$), forms radially or azimuthally polarization patterns. Low-order C-point indices, on the other hand, have three possible topologically distinct polarization configurations, which manifest as lemon or monstar for $I_C = 1/2$ ($\sigma_{12} = 1$), and star for $I_C = -1/2$ ($\sigma_{12} = -1$).

## 3. EXPERIMENT AND RESULTS

The experimental setup is depicted in Fig. 1 and includes a 4-f VOF generator system from our earlier research [22, 41] as well as a focusing component for producing a far field. The focal lengths of lenses L1, L2, and L3 are 400 mm, 300 mm, and 200 mm, respectively. We let a collimated and expanded linearly polarized laser beam of 532 nm wavelength be incident on a phase-only SLM (HOLOEYE Leto, 6.4um pixel pitch, 1920×1080), which addresses the CGH produced from the complex amplitude in Eq. (2) through a cosine-grating encoding method to generate the constituent Bessel-like modes. Then the two +1 order diffraction components are controlled to pass through a filter and converted into mutually orthogonal circularly polarized states by using an assembled quarter-wave plate (QWP). A Ronchi grating is placed at the rear focal plane of lens L2 to re-correct the diffraction direction of the two components and compose the desired Bessel-like vector beams embedded with PSs. Lens L3 is used to generate the far-field, and the unite consisting of a QWP and a polarization camera (4D TECHNOLOGY PolarCam, 3.45um pixel pitch, 2464×2056) mounted on a motorized stage can be positioned to detect and image any cross-section of the composite Bessel-like VOFs. As a result, the four stokes parameters are precisely measured, and upon which the beam patterns and the phases of the Stokes complex field $S_{12}$ will be ultimately reconstructed to verify all the predictions.

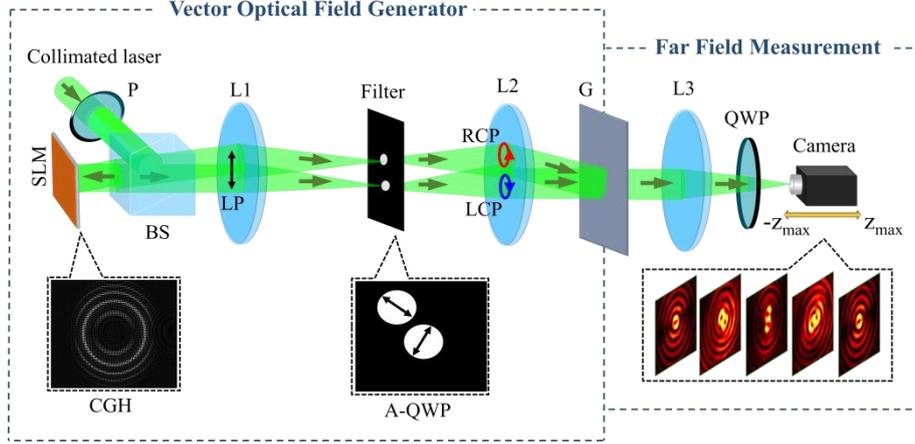

**Fig.1** Experiment setup used to generate composite Bessel-like VOFs and manipulate propagation and evolution of the embedded PSs. P, polarizer; SLM, spatial light modulator; BS, beam splitter; L1-L3, lenses; A-QWP, assembled quarter-wave plate; G, Ronchi grating.

Firstly, let us present our results regarding the manipulation of propagation trajectories of the embedded PSs in the composite field, which can be described by the displacement vector,

$$\vec{s}_i(z) = h_i(z)\hat{x} + g_i(z)\hat{y}, \quad z \in [-f, f] \quad (5)$$

where $\hat{x}$ and $\hat{y}$ denote the unit vectors of $x$ and $y$ directions, respectively. To highlight the main point, the constituent Bessel-like components are controlled to have a uniform constant intensity and each with an imposed phase vortex of unit strength, i.e. $I_L(z) = I_R(z) = 1$ and $m_L = m_R = 1$, from which it is deduced that a paired C-points, forming a lemon-star polarization topology, can be launched from any location in the Fourier plane (z = 0 mm). Then, the resultant C-lines have evolved with propagation distance as they follow pre-designed trajectories in 3D space. As an example, in Fig. 2, we numerically and experimentally demonstrate the design of the propagation along spiral trajectories, where $h_L(z) = \sin[\omega_L \pi(z/f+1)]$, $g_L(z) = \cos[\omega_L \pi(z/f+1)]$, $h_R(z) = \sin[\omega_R \pi(z/f+1)]$, and $g_R(z) = \cos[\omega_R \pi(z/f+1)]$ with $\omega_L = 3$ and $\omega_R = 4$ are set, meaning that the two C-points propagate in a similar manner but with different angular velocities. It can be seen from Figs. 2(a) and (b) that the intensity pattern features a high-intensity central core with propagation-dependent structures surrounded by numerous asymmetrical rings, which represent composite Bessel-like fields rather than exact Bessel beams due to the structured mainlobe and symmetry breakdown. The propagation trajectories of the RH and LH C-points are plotted by red and blue continuous lines (C-lines), as shown in Fig. 2(c), respectively, along with a top-view image in the forms of overlapping circles. To confirm the initial composite field with the lemon-star topological configuration, the polarization distribution in the Fourier plane is mapped as an inset, where the red and blue colors denote RH and LH polarization states, respectively, while the white and black dots mark the positive (+1/2) and negative (-1/2) C-points in x-y planes. It is found that by propagating throughout the given distance, three and four complete windings for the LH and RH C-lines are obtained, which are topologically equivalent to a PS braiding, the winding numbers of which are directly controlled by the values of $\omega_L$ and $\omega_R$ and thus can be arbitrarily regulated. Such a result is in stark contrast to the common optical vortex braiding limited to the Gouy phase [42], which may produce prospective applications in microstructure fabrication. Figs. 2(d) and (e) show the measured transverse beam patterns and phases of the Stokes field $S_{12}$ taken at various positions, with the Stokes phase vortices marked by white and black circles determining the positive and negative C-points. It can be observed that the intensity mainlobe structurally varies as the beam propagates, and the lemon-star polarization configuration rotates with respect to the central core of the beam, resulting in the braiding behavior of the C-lines. The measured locations of the Stokes vortices are found to be in good agreement with those of the C-points marked on the C-lines in Fig. 2(c).

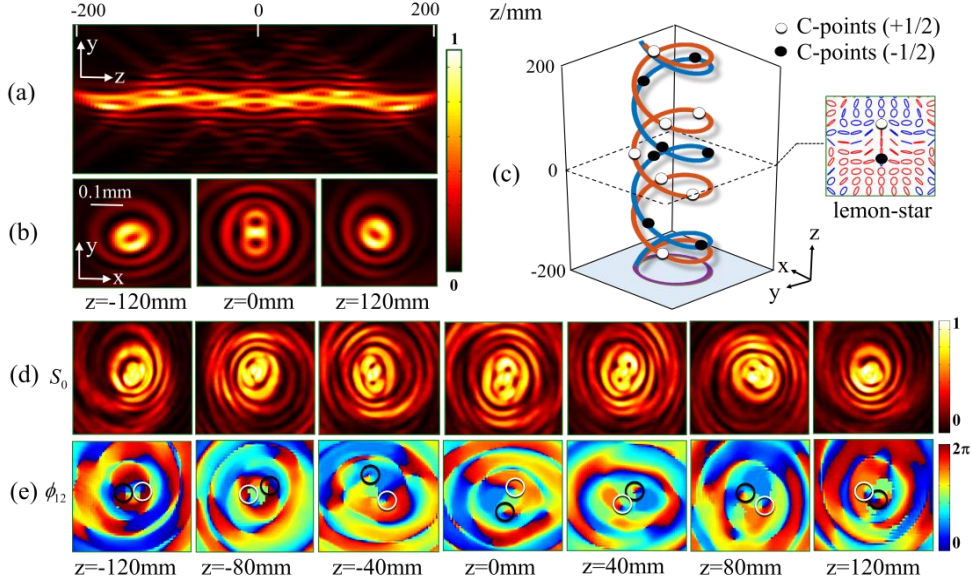

**Fig.2** Numerical and experimental demonstrations of the manipulated propagation trajectories of PSs in form of braiding. (a) Simulated side-view propagation of the composite Bessel-like beam and (b) transverse beam patterns at three different positions. (c) The spiral propagation trajectories of the C-points (i.e. C-lines, depicted by continuous lines) accompanied by a top-view image at the bottom, and the polarization distribution in the Fourier plane is indicated as an inset to suggest the initial field with lemon-star topological configuration, where the red and blue colors denote RH and LH polarization states. (d) and (e) Measured transverse beam patterns and Stokes phases $\phi_{12}$ at different positions, where the white and black circles represent the Stokes vortices, corresponding to positive and negative C-points marked by dots on the C-lines in (c), respectively.

To offer our approach more versatility, and in the meantime, prompted by the growing interest in customizing on-demand PSs topological morphology, we build an intriguing hexapetalous PSs topological configuration by separately manipulating propagation trajectories of the PSs. For this purpose, trefoil propagation trajectories of the paired C-points with the same scale but different azimuthal orientations are designed, which are described as $h_L(z)=\cos[3(z/f+1)\pi]\sin[(z/f+1)\pi]$, $g_L(z)=\cos[3(z/f+1)\pi]\cos[(z/f+1)\pi]$, $h_R(z)=\cos[3(z/f+1)\pi+3\pi/2]\sin[(z/f+1)\pi+5\pi/6]$, and $g_R(z)=\cos[3(z/f+1)\pi+3\pi/2]\cos[(z/f+1)\pi+5\pi/6]$ in Eq. (5), while the other parameters are the same as those taken in Fig. 2. The longitudinal and transverse beam patterns, propagation trajectories and their top-view images, and polarization distribution in the initial transverse plane are demonstrated in Fig. 3(a)-(c) accordingly. As expected, the C-points manifest themselves in 3D space as azimuthally dislocated trefoil propagation trajectories, as shown in Fig. 3(c), resulting in the formation of the hexapetalous PSs topological configuration in the projection plane (see the bottom plane pattern in Fig. 3(c)). Experimental results are shown in Figs. 3(d) and (e) to confirm the simulated ones presented in Figs. 3(a)-(c). In addition to the benefit of manipulating PS propagation trajectories, our proposed approach can be viewed as a flexible and easily implemented means of customizing complex PS topological configurations in 3D space.

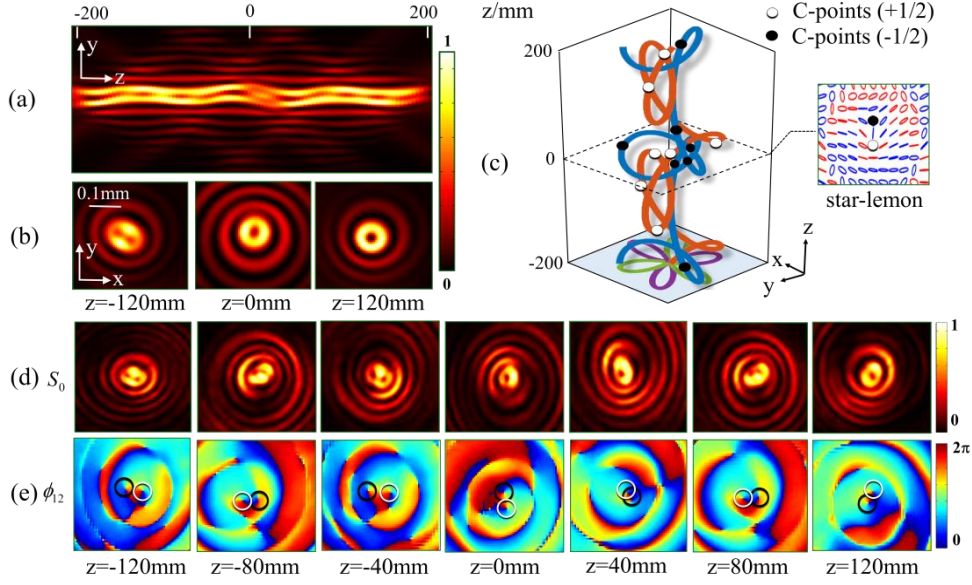

**Fig.3** Customization of the hexapetalous PSs topological configuration. (a) Simulated side-view propagation of the composite Bessel-like beam and (b) transverse beam patterns at three different positions. (c) The dislocated trefoil propagation trajectories of the C-points and their top-view images, together with an inset depicting the polarization distribution in the Fourier plane, where the red and blue colors denote RH and LH polarization states. (d) and (e) Measured transverse beam patterns and Stokes phases $\phi_{12}$ at different positions, where the Stokes vortices are marked by white and black circles, corresponding to positive and negative C-points marked by dots on the C-lines in (c), respectively.

Next, we turn our attention to the evolution of PSs during propagation. As is well known, PSs have drawn increasing attention due to their stability and persistence in time or space, which stems from the local structure organizing the topology of the surrounding field. Several novel phenomena such as PS splitting, creation, annihilation, and pseudo-topological evolution have been investigated recently [33, 43, 44], implying that the topological morphology can also be destroyed even during free-space propagation. Regardless of the strict scenario for ensuring the persistence of PS topological configurations during propagation and the underlying interaction between them, the appearance of PSs is primarily determined by orthogonal SAM states with differences in OAMs. In other words, the spatial overlap or separation of the two constituent parts determines whether PSs are generated or destroyed. Inspired by this, we attempt to regulate the embedded C-points to annihilate and revive at desired positions with beam propagation. We define $h_L(z)=0$, $h_R(z)=\sin[2\pi(z/f+1)+\pi/2]$, and $g_L(z)=g_R(z)=\cos[2\pi(z/f+1)+\pi/2]$ in Eq. (5) to ensure the same displacement of the C-points along the *y*-direction, whereas different displacements along the *x*-direction are controlled by $h_R(z)$, and the other parameters are the same as those in Fig. 2. The longitudinal and transverse beam patterns, evolution trajectories of the C-points accompanied by Stokes phase slices taken at three different positions, and the corresponding polarization distributions are shown to illustrate the evolution of the beam intensity and C-points during propagation in Fig. 4(a)-(c), where the red and blue colors represent RH and LH polarization states, respectively, and the gray dots denote the annihilation points of the positive (white dots) and negative (black dots) C-points, respectively. It is shown that, at the initial position (z=0mm), there is no PSs existed due to the overlapping of the orthogonal SAM states carrying the same OAM, i.e. $h_R(z)=h_L(z)$ and $m_L=m_R=1$. On the contrary, $h_R(z)\neq h_L(z)$ if only the beam propagates a tiny distance away from the initial position, implying the splitting of the OAM states, thus the paired C-points will be created. Furthermore, it is found that the C-points, which are caused by the pre-planned periodic displacements, alternately vanish and revive with propagating through a longitudinal distance (cf. Fig. 4 (c)). Therefore, the annihilation positions derived from the condition $h_R(z) = h_L(z)$ can be conveniently controlled. Figs. 4(d) and (e) show the measured Stokes phases and transverse beam patterns at various positions between z = -150mm and z = 150mm in steps of 50mm, where the dotted circles represent the annihilation points and the white and black circles denote the Stokes vortices, which correspond to the positive and negative C-points, respectively. It is confirmed that the C-points can be regulated to periodically annihilate at desired positions, such as z = -100 mm, z = 0 mm, and z = 100 mm. Likewise, as we can see from positions z = -150 mm and z = 50 mm, or z = -50 mm and z = 150 mm, the specific PS topological configuration can periodically revive. These results inevitably lead us to the conclusion that by manipulating the propagation trajectories of the PSs we can flexibly regulate and redistribute the constituent SAM and OAM states, allowing us to achieve the on-demand annihilation or revival of PS topological morphology in 3D space.

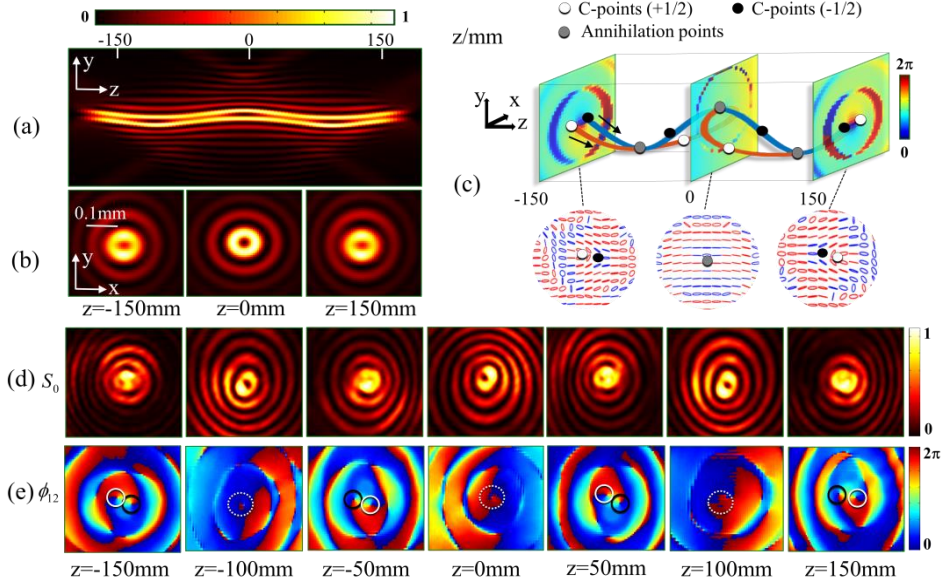

**Fig.4** Numerical and experimental demonstrations of the manipulated evolution of PSs in 3D space. (a) Simulated side-view propagation of the composite Bessel-like beam and (b) transverse beam patterns at three different positions. (c) The evolution trajectories (continuous lines) accompanied by three slices of Stokes phase and polarization distributions at the same three positions as those in (b), where the red and blue colors describe RH and LH polarization states, white and black dots denote positive and negative C-points, and gray dots represent the annihilation points of the C-points, respectively. (d) and (e) Measured transverse beam patterns and Stokes phases at different positions, where the white and black circles denote the Stokes vortices corresponding to the positive and negative C-points marked by dots on the lines in (c), and the dotted circles denote for the annihilation points.

The above examples have demonstrated the topological stability of PSs during propagation, as seen in Figs. 2-4, where a certain PS topological morphology keeps invariant along the pre-designed propagation trajectory. Such performances are favorable to information transmission fidelity when PSs are utilized as information carriers. However, another question arises: can we switch the PS structure of an optical field during its propagation? The answer is affirmative and is supported by the final example. To this end, we carry out a piecewise design, which integrates multiple component phase vortices ($m_L, m_R$) = (+1, +1), (+1, -1), (+1, +1), and (-1, +1), within their respective z ranges (-200mm, -150mm), (-150mm, -50mm), (-50mm, 50mm), and (50mm, 150mm). The displacement equations are set to be $h_R(z) = h_L(z) = 0$ and $g_{L,R}(z) = \pm\sin[2\pi(z/f+1)+\pi/2]$ so that during propagation, two C-points displace oppositely about the optical axis solely in the y direction. The simulated and experimentally measured results are demonstrated in Fig. 5, where it is first shown that a chain-like intensity pattern in the y-z plane is formed (Fig. 5 (a)), and the transverse beam patterns at three positions are presented in Fig. 5(b) to further illustrate such an intensity distribution. The evolution trajectories of PSs with three Stokes phase slices are shown in Fig. 5(c), and the polarization distributions at six typical places are shown as insets to highlight the variation of topological morphology. It is found that, in regions (-150mm, -50mm), (-50mm, 50mm), and (50mm, 150mm), the composite beam displays three different PS topological configurations, namely, lemon-lemon, star-lemon, and star-star, respectively, resulting from the piecewise design of constituent phase vortices. Moreover, at the overlapping points, the PSs can not only annihilate (gray dots), but can also be fused into different higher-order modes, e.g., at position z=-50mm, the paired positive C-points (white dots) are transformed into a positive V-point (green dot), around which the beam is radially and linearly polarized. Whereas, a negative V-point (yellow point) with a spider-like linear polarization configuration appears at position z=150mm, due to that the signs of the paired C-points are changed to be negative before the fusion point. Figs. 5 (d) and (e) show the measured transverse beam patterns and phases of Stokes field $S_{12}$ at different positions, where the white, black, green, and yellow circles represent the Stokes vortices corresponding to the positive (+1/2) C-points, negative (-1/2) C-points, positive (+1) V-point, and negative (-1) V-point, marked on the lines in (c), respectively, while the dotted circles reflect the annihilation of PSs. Obviously, the piecewise design of PS topological configurations, annihilation and transformation of C-points, as well as the periodic variations of transverse intensity corresponding to the longitudinal chain-like pattern, are completely confirmed.

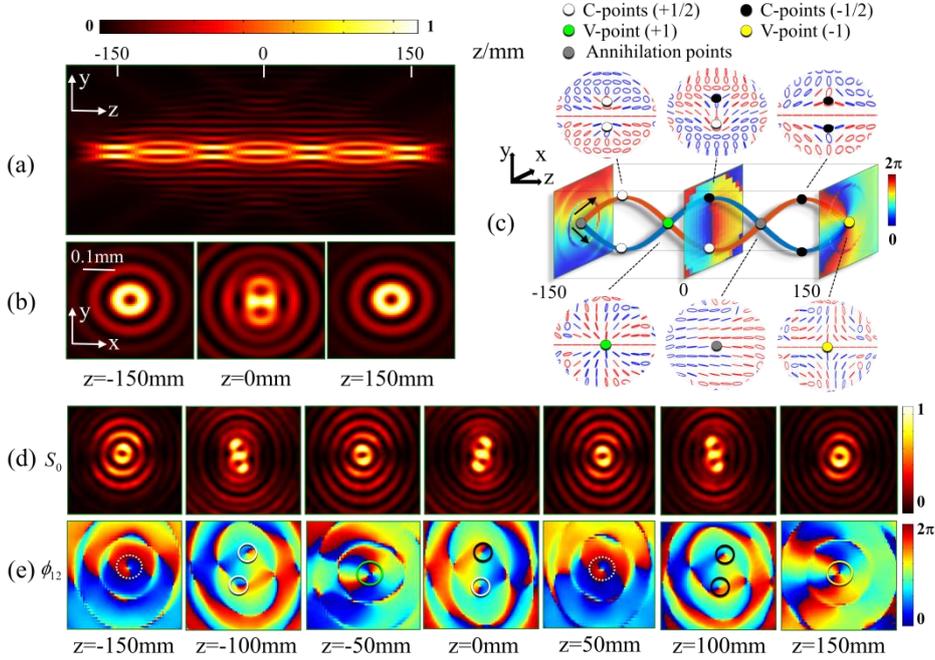

**Fig.5** The piecewise design of the z-axis evolution of PS topological morphology. (a) Simulated side-view propagation of the composite Bessel-like beam and (b) transverse beam patterns at three typical positions. (c) The evolution trajectories (continuous lines) with three slices of Stokes phase and polarization distributions at six characteristic positions, where the red and blue colors describe RH and LH polarization states, white

and black dots denote positive and negative C-points, green and yellow dots denote positive and negative V-points, and gray dots represent the annihilation points of PSs, respectively. (d) and (e) Measured transverse beam patterns and Stokes phases at different positions. In (e), the white, black, green, and yellow circles represent the Stokes vortices corresponding to the positive C-points, negative C-points, positive V-point, and negative V-point, respectively, while the dotted circles denote the annihilation points.

## 4. CONCLUSION AND DISCUSSION

In summary, we have presented a convenient approach to manipulate the propagation and evolution of PSs in composite Bessel-like VOFs. Based on our experiment setup comprising a VOF generator and a far-field measurement system, we generated the initial Bessel-like beams embedded with PSs and performed a thorough investigation on how the propagation trajectories of the PSs are flexibly manipulated by our proposed approach. We first verified the feasibility of the approach by designing a PS braiding, whose winding numbers can be arbitrarily controlled. In addition, an interesting hexapetalous PSs topological configuration is customized by prescribing two dislocated trefoil propagation trajectories to the paired C-points to display the versatility of the approach. Second, we highlighted the importance of the persistence of PS topological morphology during the propagation process, elucidated the physical mechanism that the formation or destruction of a specific PS is resulted from the overlapping or separation of the constituent SAM and OAM states, and further demonstrated how to regulate the evolution of PSs, e.g. annihilation and revival at desired positions with tunable longitudinal periods. Finally, to break though the persistence of PS topology upon propagation and manipulate the evolution with multi-degree of freedom, we carry out a piecewise design from several combinations of component phase vortices to arbitrarily regulate the PS topological morphology, annihilation, and transformation between them. It should be stressed that while 3D PS topological morphology types like knots, links, and Möbius strips directly reflect the spatial evolution of the complex PSs, controlling PS propagation and evolution and customizing desired 3D PS topological morphology are still challenging and desearve more research efforts. Our approach provides a new way for harnessing and exploiting PSs and may facilitate applications of singular optics in the fields like information processing and optical communication.

**Funding.** National Key R&D Program of China (2018YFA0306200, 2017YFA0303700); National Natural Science Foundation of China (NCFC) (91750202, 11922406); Training Program of Anhui Polytechnic University (S022020077, 2016yyzr11).

**Acknowledgments.**

## REFERENCES

1. J. Ruostekoski and Z. Dutton, "Engineering vortex rings and systems for controlled studies of vortex interactions in Bose-Einstein condensates," Phys. Rev. A **72**, 063626(2005).
2. T. Isoshima, "Vortex chain structure in Bose-Einstein condensates," J. Phys. Soc. Jpn. **77**, 094001(2008).
3. H. K. Moffatt, "The degree of knottedness of tangled vortex lines (vol 35, pg 117, 1969 )," J. Fluid Mech. **830**, 821-822(2017).
4. A. A. Abrikosov, "On the magnetic properties of superconductors of the second group," Sov. Phys. JETP **5**, 1174-1183(1957).
5. L. Faddeev and A. J. Niemi, "Stable knot-like structures in classical field theory," Nature **387**, 58-61(1997).
6. R. A. Battye and P. M. Sutcliffe, "Knots as stable soliton solutions in a three-dimensional classical field theory," Phys. Rev. Lett. **81**, 4798-4801(1998).
7. J. F. Nye and M. V. Berry, "Dislocations in wave trains," Proc. R. Soc. London Ser. A **336**, 165-190(1974).

8. J. F. Nye and P. Sciences, "Lines of circular polarization in electromagnetic wave fields," Proc. R. Soc. London Ser. A **389**, 279-290(1983).
9. I. I. Freund and N. Shvartsman, "Wave-field phase singularities: The sign principle," Phys. Rev. A **50**, 5164-5172(1994).
10. M. R. Dennis, "Polarization singularities in paraxial vector fields: morphology and statistics," Opt. Commun. **213**, 201-221 (2002).
11. I. Freund, "Polarization singularity indices in Gaussian laser beams," Opt. Commun. **201**, 251-270(2002).
12. F. Flossmann, K. O'Holleran, M. R. Dennis, and M. J. Padgett, "Polarization singularities in 2D and 3D speckle fields," Phys. Rev. Lett. **100**, 203902(2008).
13. F. Flossmann, U. T. Schwarz, M. Maier, and M. R. Dennis, "Polarization singularities from unfolding an optical vortex through a birefringent crystal," Phys. Rev. Lett. **95**, 253901(2005).
14. I. Freund, "Cones, spirals, and Möbius strips, in elliptically polarized light," Opt. Commun. **249**, 7-22 (2005).
15. H. Larocque, D. Sugic, D. Mortimer, A. J. Taylor, R. Fickler, R. W. Boyd, M. R. Dennis, and E. Karimi, "Reconstructing the topology of optical polarization knots," Nat. Phys. **14**, 1079-1082(2018).
16. F. Maucher, S. Skupin, S. A. Gardiner, and I. G. Hughes, "Creating Complex Optical Longitudinal Polarization Structures," Phys. Rev. Lett. **120**, 163903(2018).
17. D. Sugic and M. R. Dennis, "Singular knot bundle in light," J. Opt. Soc. Am. A **35**, 1987-1999(2018).
18. Y. Shen, X. Wang, Z. Xie, C. Min, X. Fu, Q. Liu, M. Gong, and X. Yuan, "Optical vortices 30 years on: OAM manipulation from topological charge to multiple singularities," Light Sci Appl. **8**, 90(2019).
19. E. Otte and C. Denz, "Customization and analysis of structured singular light fields," J. Opt. **23**, 073501(2021).
20. Q. Wang, C.-H. Tu, Y.-N. Li, and H.-T. Wang, "Polarization singularities: Progress, fundamental physics, and prospects," APL Photonics **6**, 040901(2021).
21. F. Cardano, E. Karimi, L. Marrucci, C. de Lisio, and E. Santamato, "Generation and dynamics of optical beams with polarization singularities," Opt. Express **21**, 8815-8820(2013).
22. X.-L. Wang, J. Ding, W.-J. Ni, C.-S. Guo, and H.-T. Wang, "Generation of arbitrary vector beams with a spatial light modulator and a common path interferometric arrangement," Opt. Lett. **32**, 3549-3551(2007).
23. S. W. D. Lim, J. S. Park, M. L. Meretska, A. H. Dorrah, and F. Capasso, "Engineering phase and polarization singularity sheets," Nat. Commun. **12**, 4190(2021).
24. I. J. O. l. Freund, "Polarization singularities in optical lattices," Opt. Lett. **29**, 875-877(2004).
25. P. Kurzynowski, W. A. Wozniak, M. Zdunek, and M. Borwinska, "Singularities of interference of three waves with different polarization states," Opt. Express **20**, 26755-26765(2012).
26. R. Yu, Y. Xin, Q. Zhao, Y. Chen, and Q. Song, "Array of polarization singularities in interference of three waves," J. Opt. Soc. Am. A **30**, 2556-2560(2013).
27. S. K. Pal, G. Arora, Ruchi, and P. Senthilkumaran, "Handedness control in polarization lattice fields by using spiral phase filters, " Appl. Phys. Lett. **119**, 221106(2021).
28. S. K. Pal and P. Senthilkumaran, "Lattice of C points at intensity nulls," Opt. Lett. **43**, 1259-1262(2018).
29. C. Chang, L. Li, Y. Gao, S. Nie, Z.-C. Ren, J. Ding, and H.-T. Wang, "Tunable polarization singularity array enabled using superposition of vector curvilinear beams," Appl. Phys. Lett. **114**, 041101(2019).
30. X. Wang, Z. Li, Y. Gao, Z. Yuan, W. Yan, Z.-C. Ren, X.-L. Wang, J. Ding, and H.-T. Wang, "Configuring Polarization Singularity Array Composed of C-Point Pairs," IEEE Photon. J. **14**(4), 1-6(2022)
31. T. H. Lu, Y. F. Chen, and K. F. Huang, "Generalized hyperboloid structures of polarization singularities in Laguerre-Gaussian vector fields," Phys. Rev. A **76**, 063809(2007).
32. Y. Wang and G. Gbur, "Hilbert's Hotel in polarization singularities," Opt. Lett. **42**, 5154-5157 (2017).

33. G. L. Zhang, M. Q. Cai, X. L. He, X. Z. Gao, M. D. Zhao, D. Wang, Y. Li, C. Tu, and H. T. Wangrmark, "Pseudo-topological property of Julia fractal vector optical fields," Opt. Express **27**, 13263-13279(2019).
34. R. Droop, E. Otte, and C. Denz, "Self-imaging vectorial singularity networks in 3d structured light fields," J.Opt. **23**, 074003(2021).
35. W. Yan, Y. Gao, Z. Yuan, Z. Wang, Z. C. Ren, X. L. Wang, J. Ding, and H. T. Wang, "Non-diffracting and self-accelerating Bessel beams with on-demand tailored intensity profiles along arbitrary trajectories," Opt. Lett. **46**, 1494-1497(2021).
36. T. Cizmar and K. Dholakia, "Tunable Bessel light modes: engineering the axial propagation," Opt. Express **17**, 15558-15570(2009).
37. P. Li, Y. Zhang, S. Liu, L. Han, H. Cheng, F. Yu, and J. Zhao, "Quasi-Bessel beams with longitudinally varying polarization state generated by employing spectrum engineering," Opt. Lett. **41**, 4811-4814(2016).
38. V. Jarutis, A. Matijosius, P. Di Trapani, and A. Piskarskas, "Spiraling zero-order Bessel beam," Opt. Lett. **34**, 2129-2131 (2009).
39. J. Zhao, P. Zhang, D. Deng, J. Liu, Y. Gao, I. D. Chremmos, N. K. Efremidis, D. N. Christodoulides, and Z. Chen, "Observation of self-accelerating Bessel-like optical beams along arbitrary trajectories," Opt. Lett. **38**, 498-500(2013).
40. C. Liang, Z. Yuan, W. Yan, Y. Gao, X. Wang, Z. C. Ren, X. L. Wang, J. Ding, and H. T. Wang, "Radially self-accelerating Stokes vortices in nondiffracting Bessel-Poincare beams," Appl. Opt. **60**, 8659-8666(2021).
41. C. Chang, Y. Gao, J. Xia, S. Nie, and J. Ding, "Shaping of optical vector beams in three dimensions," Opt. Lett. **42**, 3884-3887 (2017).
42. A. A. Voitiv, J. M. Andersen, M. E. Siemens, and M. T. Lusk, "Optical vortex braiding with Bessel beams," Opt. Lett. **45**, 1321-1324(2020).
43. S. Vyas, Y. Kozawa, and S. Sato, "Polarization singularities in superposition of vector beams," Opt. Express **21**, 8972-8986(2013).
44. E. Otte, C. Alpmann, and C. Denz, "Polarization Singularity Explosions in Tailored Light Fields," Laser Photonics Rev. **12**, 1700200(2018).